# Sound Conveyors for Stealthy Data Transmission

Sachith Dassanayaka

Department of Mathematics & Computer Science, Wittenberg University, Ohio, 45501, USA
dassanayakas@wittenberg.edu

**Abstract** – Hiding messages for countless security purposes has become a highly fascinating subject nowadays. Encryption facilitates the data hiding. With the express development of technology, people tend to figure out a method capable of hiding a message and the survival of the message. The present-day study is conducted to hide information in an audio file. Generally, steganography advantages are not used among industry and learners even though it is an extensively discussed area in the present information world. This implementation aims to hide a document such as .txt, .doc, and .pdf file formats in an audio file and retrieve the hidden document when necessary. This system is called DeepAudio v1.0. The system supports AES encryption and tolerates both wave and MP3 files. The sub-aims of this work were the creation of a free, openly available, bug-free software tool with additional features that are new to the area.

## 1. Introduction

The rapid development of information and communication technologies is constantly increasing, and demands are being made to achieve maximum safety and reliability in these areas. This is a current and highly debated issue, and there is evidence that such a situation will persist. Currently, there are many methods and disciplines dealing with the provision of information and communication systems. During the 'information age,' information has become very important as it implies some way to protect it [1]. Of course, the protection or secrecy of the information is not new. People of ancient Greece first mentioned attempts to classify essential messages that could affect the course of war. The importance of confidentiality of sensitive information in the military sector has persisted to this day, but it is very important in other areas, mainly due to the rapid development and expansion of information technology and computer networks. Sometimes, we do not need the information to protect confidentiality, but on the contrary, we want to be free to expand while maintaining our proof of authorship. These programs are the most frequently used image file formats, such as cover (sometimes called carrier).

Just as steganography conceals messages in seemingly innocuous files, fake reviews may contain hidden meanings or encoded messages that are not immediately apparent to casual readers. Some advanced fake review generators might use linguistic steganography techniques to embed hidden information within the text of reviews, making them appear genuine while conveying additional data. Fake review generators might adopt more sophisticated steganographic methods to evade detection [2].

### 1.1 Motivation

Steganography of audio signals is more challenging than the steganography of images or video sequences due to the wider energetic range of the Human auditory system compared to the Human visual system. Only a few algorithms have been developed to embed a message into an audio file. The surviving systems require a considerable time to embed a small message [3]. Furthermore, we have found that the local people/authorities are not aware of steganography, even though it is widely discussed in the modern information world. This research is an extension of the previously developed AudiSteg v1.0 application [4].

### 1.2 Aims and Objectives

Our goal was to create a software tool for hiding valuable files within an audio file. Therefore, it was necessary to revise the proposed program to extract the concealed data directly from the audio. This process has collaborated to follow ordered objectives.
- ➢ Enhance cryptographic security of hidden data (Confidentiality) and passwords (Authentication).
- ➢ Increase the rate of embedded data.
- ➢ Assure exactness of decrypted data (Integrity).
- ➢ Assure unapparent perceptual transparency of the audio file (Cover object) and the object containing secret messages.

- ➢ Perfectly sending audio files to another location or party through any standard network.
- ➢ Prove the quality of the developed program by analyzing the survey results.

## 2. Literature Review

Steganography can be defined as "The art of hiding and transmitting data through apparently innocuous transporters in an effort to cover the existence of the data." Steganography is a Greek-origin word that means "hidden writing." The word can be divided into two parts: Steganos and Graphic. Stegnos means "Secret" or "covered." And graphic means "Writing." The book "Steganographia," written by Johannes Trithemius in 1499, was the first book covering steganography techniques [3].

In general, many operating systems use the basics of steganography to hide files and folders. In Windows OS, we can hide a folder or a file in another folder using Hidden Directories or in UNIX hiding directories. Covert channels are used in networks to transfer valuable data in apparently usual network traffic. Furthermore, encryption software called TrueCrypt also uses steganography to secure hidden data [5]. With that, they can use different passwords to hide a hidden TrueCrypt volume or a hidden operating system within the main outer TrueCrypt volume. Nobody can easily find the existence of hidden data volume or the hidden operating system even though he knows that it's a TrueCrypt volume and has the outer volume password.

### 2.1 Steganography Types

Two prominent steganography types exist [6].
- Pure Steganography- Hide valuable information inside the carrier file without an additional shield.
- Secret key Steganography- Hide valuable information inside the carrier files with an additional shield (Password).

Cryptography and steganography seem to be similar to each other. However, some similarities and deviations can be summarized as follows.

TABLE I
Comparison of Cryptography vs. Steganography

| Cryptography | Steganography |
|---|---|
| Similarities | |
| Uses for Information Security. | Uses for Information Security. |
| Uses Symmetric and Asymmetric keys. | Uses Symmetric Keys for Embedding and Extracting. |
| Differences | |
| Mainly Concerns on Protecting information. | Mainly concerned with concealing the existence of hidden information. |
| A third party can notice cryptography. | Hide the evidence of secret information. |
| Common technology. All know most algorithms. | Less known technology. Algorithms are developing |

## 3. Theory and Methodology

The prominent logic of this research is based on specific theories in computer science as well as statistics.

### 3.1 Sampling Rate

Capture audio covering the entire 20–20,000 Hz range of human hearing, such as when recording music or many types of acoustic events, and audio waveforms are typically sampled at 44.1 kHz, 48 kHz, 88.2 kHz, or 96 kHz. The approximately double-rate condition is a consequence of the Nyquist theorem. Sampling rates higher than 50 kHz to 60 kHz cannot supply more usable information for human listeners [7].

### 3.2 Advance Encryption Standard

AES is not a Feistel cipher. It works in parallel over the whole input block. Designed to be efficient in hardware and software through various platforms. Symmetric or secret-key ciphers use the same key for encryption and decryption, so the receiver must know the same key to decrypt the message. Block size is 128-bit (but also 192 or 256-bit). Key length can be 128, 192, or 256 bits. Number of rounds is 10, 12 or 14. Key scheduling is 44, 52, or 60 subkeys with length = 32 bit [8].

### 3.4 Least significant bit

LSB is the most conjoint steganography technique used in real-world scenarios for image cover objects [9]. The theory says that the modification should occur at the last bit of the particular byte stream.

## 4. Proposed way-out

The resolution includes different approaches, but each style performs on the same generic architecture.

### 4.1 LSB Encoding

The system has developed on the top of three distinct logics. According to the selected logic and options, the system will present the maximum data size that can be hidden before executing the process. The first approach belongs to the least significant bit (LSB) technique, and the system that uses the LSB technique is called Method I. Though Method I is a complex process, it is flexible. In the first place, the system copies an audio file to buffer and, at the same time, calculates the size of the document and writes that value inside 4 bytes using an 8-bit right shift [10]. Skip the audio header and then write the document size next to the audio header. Then, read the message (UTF-16) or document and encrypt using AES 256 with a given key. Then, start to write the encrypted message inside the audio from the last used position. If a user requires compression to escalate the size of the message, then change the LSB for each byte. Otherwise, skip the first 8 bits and change LSB in the next 8bit until the end of the message. The system changes bits from the left and right channels of each sample. It reduces the noise inside the audio during modifications. At the end of the process, encoded audio is written in a given location, like the original audio file.

The retrieval process requires encoded audio and the secret key to encode the message or document. Read the size and other information of hidden data and retrieve the correct bits from the audio byte stream. Use left shifting and calculate message size from retrieved bytes. Then, decrypt the extracted message and save the document in the given location. Users can remove the message from audio after it is extracted if required. The encode or message removal process does not affect the audio file.

### 4.2 Injection Encoding

Injection is a quite simple method that directly injects the secret information into the carrier file. The payload and carrier message is directly fed into the specially designed stegosystem encoder. Wherever the proposed system refers, this technique is called method II. Each audio file consists of a header and body, including summarized details about the audio and data [11]. This technique inserts the summarized message information according to the message in the middle of the header and body and then attaches the secret message or document. Method II is a simple process but it is very powerful and much quicker than the previous technique. The system copies audio bits to buffer and calculates the document or text message size. Write the header and then add five bytes representing the size of hidden data and other user-select options, such as compression details. Then, write the rest of the audio into the buffer. Then, it reads the message and encrypts using AES 256 with a given key. If a user requires compressing the message, then it processes before encrypting. Write the encrypted and compressed message at the end of the buffer. Finally, writes the entire encoded audio in the given location.

The retrieval process of Method II skips the header bytes of an audio file and reads the next five bytes. Then, it points to the end of the audio file and retrieves the hidden message. Next, decrypt the message and decompress. Finally, the document is saved in the user's given location.

### 4.3 MP3 Encoding

The MP3 steganography process involves manipulating the file's packet structure, which consists of headers and data fields. The encoding begins by skipping the header of the first packet and writing the encrypted message size in its data field. Subsequently, the message is embedded into the data fields of subsequent packets using the Least Significant Bit (LSB) technique [12]. A constant skip value is employed between modified packets to maintain audio quality, with higher skip values generally resulting in better sound quality. The encoded MP3 file is then saved to a user-specified location. Due to size constraints, only text files or messages are allowed, with a maximum size limit of 1/16 (6.25%) of the MP3 carrier file size. If this limit is exceeded, the system will not proceed and will notify the user. The retrieval process reverses these steps: it checks the first packet for the message size, reads the message bits from the data fields using the predefined skip value, decrypts the extracted bits, and saves the result as a text file. Notably, preserving the original quality of the MP3 file while embedding data remains an extremely challenging aspect of this process.

### 4.4 Send and Receive Encoded File

The system is designed to facilitate the transfer of audio files within a Local Area Network (LAN) environment, specifically tailored for users of DeepAudio v1.0. It allows senders to transmit encoded audio files to target receivers, with these files being treated as normal audio files within the system. To initiate a transfer, the sender must input the receiver's IP address, and the connection is only established once the receiver accepts the file transfer request. Each sender-receiver pair operates on a single connection through a randomly assigned socket. The system supports multiple simultaneous file transfers without imposing size restrictions, and the file transfer process operates independently from the encoding and decoding processes. This design ensures efficient and secure file sharing while maintaining the integrity of the steganographic features of DeepAudio v1.0.

## 5. Testing

The following test results are based on a 25.4 MB audio (wave) file, which is the same size as .txt, .doc, and .pdf document formats. The test cycle has been completed 60 times per single message size and 20 times per document format. However, according to the above-proposed technique, the consequences have been displayed as an average time for the above three types of document formats. The average time for both encoding and decoding has been summarized in Table II.

### 5.1 Noise

While editing the audio files, the noise inside the audio file becomes a prominent fact, and the noise increases according to the edited portion in a particular audio file. The noise inside an audio file increases according to the size of the hidden text/document, but this will only affect Method I because Method II does not edit the inner part of the audio. However, the noise change is hard to detect using software because it is negligible, and the system has restricted the size of hidden data, which is a maximum of 1/8 from its carrier for Method I to minimize the noise increment as well.

TABLE II
Average time required for both encoding and decoding according to proposed way-outs.

| | Method I (LSB) | | Method II (Injection) | |
|---|---|---|---|---|
| Message (kb) | Encode time (s) | Decode time (s) | Encode time(s) | Decode time (s) |
| 2.2 | 0.01 | 1.29 | 0.10 | 0.10 |
| 20.1 | 0.10 | 8.61 | 0.13 | 0.11 |
| 46 | 0.18 | 12.44 | 0.20 | 0.12 |
| 59.6 | 0.20 | 15.83 | 0.22 | 0.13 |
| 110 | 0.34 | 33.96 | 0.30 | 0.18 |
| 207.5 | 0.42 | 71.76 | 0.36 | 0.20 |
| 240 | 0.44 | 82.07 | 0.41 | 0.22 |
| 1047.8 | 0.50 | 306.54 | 0.58 | 0.29 |
| 2095.8 | 1.1 | 682.32 | 0.91 | 0.35 |

### 5.2 Quality Assurance

A smoke test has been used to evaluate the quality and performance of the system (see Table III).

TABLE III
Smoke test to evaluate the quality and performance of the system

| Test case | Result | Comments |
|---|---|---|
| Users are able to get an idea without having documents | Pass | The main interface has displayed a description. |
| User-friendly GUI | Pass | Separate fields have distinct color |
| Users are encouraged to give feedback. | Pass | Developer email and information have been presented. |
| Platform independent software | Fail | Supports only Windows OS |
| Computer performance is not important when running software. | Pass | Requires small processing power |
| Virus-free software | Pass | |
| Different scenarios are provided to do the same task | Pass | Few techniques exist |
| Supports all kinds of audio formats | Fail | Wave and MP3 only |
| Allows all kinds of file types to hide | Fail | .txt, .doc, and. .pdf formats only. |
| Users can add audio or hide messages from any location | Pass | |
| Users can Save encoded or decoded files to any location | Pass | |
| Encryption methods are available. | Pass | Powered by AES 256 encryption. |
| Users can delete the message without affecting the audio file. | Pass | Occurs only if the user confirms the action. |
| Each result is shown to the user by using messages. | Pass | Keeps user interaction. |
| The software allows running in the background | Pass | It allows for minimizing and triggering messages when the process is done. |
| Any file format can be sent or received via the software | Fail | Only wave and MP3 files. |
| Exactness of message before and after the process. | Pass | Both messages are the same |
| Exactness of Audio before and after the process | Pass | Both files have the same properties, such as sample rate and music. However, only Method I keeps the same size due to the technique. |

The exactness of the original message and message after retrieval have been tested Using open-source software called Winmerge version 2.14.0. WinMerge is a freeware, differencing, and integration tool for Windows. This software can compare both folders and files, giving differences in a text format that is easy to comprehend and handle [13].

## 6. Performance

A survey was conducted to measure the system performance among professionals and identify the system's quality, which satisfies the expectations of IT professionals. Acceptance of the system's availability on the Internet is a prominent area that has been considered.

### 6.1 Explore Results

Table IV, illustrates the analyzed results of program quality and wishes for this program to be available on the Internet.

TABLE IV
Summarized results of prominent areas in the survey.

| Subject | Quality of the program | | |
|---|---|---|---|
| | Yes | No | No Idea |
| Exactness of secret message before and after the process | 96.21% | 2.13% | 1.66% |
| The exactness of audio before and after the process | 86.31% | 12.68% | 1.01% |
| Availability in the Internet | 89.59% | 6.03% | 4.38% |

A survey on the effectiveness of DeepAudio v1.0 revealed that 96.21% of users confirmed the retrieved message was identical to the original, while 2.13% noted some discrepancies. Regarding the carrier's integrity, 86.31% believed the program maintained the original quality, but 12.68% reported issues. A small percentage (1.66% and 1.01%) were uncertain about message and carrier integrity, respectively. The system's potential availability on the Internet was another crucial aspect evaluated. A significant majority (89.59%) approved of the program's quality and supported its online availability, while 6.03% expressed negative opinions. Some concerns were raised about the potential misuse of DeepAudio v1.0 for spreading misinformation through audio carriers[14, 15]. This feedback highlights the overall positive reception of the software's effectiveness and potential while acknowledging the need to address concerns about its responsible use.

## 7. Discussion

The current study was done to hide a message in an audio. The importance of hiding in an audio is sheltered because it less susceptible. It can reduce the man-in-the-middle attack. The main technique is to change the LSB value of the carrier file. In order to do that, the audio is separated into a set of byte values and messages as well. The LSB value in the carrier file is changed in each byte according to the message bytes so that the LSB carries the hidden message. In addition, the injection technique hides text messages in an audio file. It gives accurate and decent security for data hiding. It also provides a cryptographic way to secure audio file data. Suppose that someone could separate message bytes from the original audio file. If that person needs to read the secret message first, he should break the AES encryption because the hidden message was encrypted from AES-256. However, the system ensures the original quality of the carrier after encoding and the exactness of the hidden information after retrieving it.

### 7.1 Comparison

Although many free arrangements hide information inside images, only a few free surviving systems allow the user to hide information inside audio files. This comparison considers only the differences between existing systems and DeepAudio v1.0.

- Powered by AES-256 encryption.
- Separated collections.
- Supports different types of documents to hide, such as pdf, doc, and txt.
- More user-friendly GUI.
- The program will never get stuck during encode or decode process.
- Hidden messages can be destroyed without affecting the carrier.
- Allows sending and Receiving audio files via LAN.
- Processes can be run simultaneously.

### 7.2 Limitations

LSB (Method I) - The basic limitation is the size of the message. Theoretically, the LSB was changed, and the message should be 8 times smaller than the audio file. But, technically, the message size is not exactly the same and also less than the theoretical value. Audio files are very sensitive, so it is difficult to change each byte in the audio stream. According to the size of an audio file, encode and decode time increases.

Injection - The most common issue is increasing encoded file size. (Original audio + message)

MP3 (Method I) - Mp3 files are lossy file formats [14]. They do not have enough space to hide big messages. Furthermore, MP3 files are more difficult to change than wave files because they are already compressed and remove specific qualities from the audio. So, only text files or text messages can be hidden, and the size is restricted to 1/16 from its carrier as well.

## 8. Conclusion

The primary objective of this research was to develop a system capable of concealing messages or files within audio carriers, a process known as audio steganography. The system, named DeepAudio v1.0, was rigorously evaluated for its core functionalities, particularly its ability to hide messages and accurately extract them when needed effectively. The results of these evaluations were overwhelmingly positive, demonstrating the system's proficiency in both concealment and retrieval of hidden data.

DeepAudio v1.0 employs two main techniques for data hiding. The first is the Least Significant Bit (LSB) technique, which involves altering the least significant bit of the audio file's data according to the bits of the text message. This method is notable for its subtlety, as the size difference between the original and encoded audio files is virtually undetectable. The message is first converted into byte code before being embedded into the carrier file, with the message size cleverly written into the audio file header. The second technique, known as the injection technique, offers even greater capacity for hidden data, allowing for larger messages to be concealed without significant limitations.

Both approaches in DeepAudio v1.0 utilize a compression technique with three levels: Low, Medium, and High, with Medium being the default setting. This compression not only allows for more information to be hidden but also ensures the quality of the concealed message. The system implements the Advanced Encryption Standard (AES) cryptography to enhance security further. This additional layer of protection means that even if someone were to extract the hidden data from the audio file successfully, they would still need to break the encryption to access the actual message. Moreover, the system supports two-tier protection using cryptographic associations, significantly bolstering its security features.

DeepAudio v1.0 goes beyond mere steganography by incorporating file transfer capabilities. Users can send and receive audio files from one computer to another within a network, enhancing the system's utility and user interaction. A comprehensive user manual is provided to facilitate easy usage, guiding users through the application's features and functionalities. The system has been rigorously tested to handle exceptions gracefully, displaying user-friendly messages when issues arise.

The combination of cryptography and steganography in DeepAudio v1.0 results in a high degree of security, setting it apart from many other available software tools in the field of information security. When compared to existing systems, DeepAudio v1.0 offers enhanced features, improved security, and faster performance. It strikes a balance between security, message capacity, robustness, and speed, making it a formidable competitor in the realm of steganographic tools.

The effectiveness and potential of DeepAudio v1.0 are further validated by survey results, which indicate that over 89% of IT professionals who tested the application approve of it and would like to see it available on the Internet. This overwhelmingly positive reception suggests that DeepAudio v1.0 not only met but exceeded its initial research goals, positioning it as a promising tool in the field of audio steganography and secure communication.

# Declarations

## Authors' contributions

Sachith E. Dassanayaka contributed to the experiment design, performed the data, and drafted the manuscript.

# References

[1]. Castells, M. (2011). The rise of the network society. John wiley & sons.

[2]. Jayasinghe, J. T., & Dassanayaka, S. (2024). Detecting Deception: Employing Deep Neural Networks for Fraudulent Review Detection on Amazon. https://doi.org/10.21203/rs.3.rs-5214171/v1

[3]. Simmons, G.J. (1984). The Prisoners' Problem and the Subliminal Channel. In: Chaum, D. (eds) Advances in Cryptology. Springer, Boston, MA. https://doi.org/10.1007/978-1-4684-4730-9_5.

[4] D. M. S. Eranga and H. D. Weerasinghe, "Audio transporters for unrevealed communication," 2015 Fifteenth International Conference on Advances in ICT for Emerging Regions (ICTer), Colombo, Sri Lanka, 2015, pp. 267-267, doi: 10.1109/ICTER.2015.7377700.

[5]. Walter Bender, Daniel Gruhl, Norishige Morimoto, Anthony Lu, Techniques for Data Hiding, vol.35, no. 3 and 4, (IBM Systems Journal,1996), p. 313-336.

[6]. Soumyendu Das, Subhendu Das, Bijoy Bandyopadhyay and Sugata Sanyal, Steganography and Steganalysis: Different Approaches, Vol. 2, No 1, (International Journal of Computers, Information Technology and Engineering (IJCITAE), Serial Publications, 2008).

[7]. Shannon, C. E., Communication in the presence of noise, vol. 37, no. 1, (Proc. Institute of Radio Engineers, 1949), pp. 10–21.

[8] Jamil, T. (2004). The rijndael algorithm. IEEE potentials, 23(2), 36-38.

[9]. Djebbar, F., Ayad, B., Hamam, H., & Abed-Meraim, K. (2011, April). A view on latest audio steganography techniques. In 2011 International Conference on Innovations in Information Technology (pp. 409-414). IEEE.

[10]. Cvejic, N., & Seppanen, T. (2002, December). Increasing the capacity of LSB-based audio steganography. In 2002 IEEE Workshop on Multimedia Signal Processing. (pp. 336-338). IEEE.

[11]. Fleischman, E. (1998). WAVE and AVI codec registries (No. rfc2361).

[12]. Brandenburg, K. (1999, September). MP3 and AAC explained. In Audio Engineering Society Conference: 17th International Conference: High-Quality Audio Coding. Audio Engineering Society.

[13]. Marcus, A., & Poshyvanyk, D. (2005, September). The conceptual cohesion of classes. In 21st IEEE International Conference on Software Maintenance (ICSM'05) (pp. 133-142). IEEE.

[14]. Swed, O., Dassanayaka, S., Volchenkov, D.:Keeping it authentic: the social footprint ofthe trolls' network. Social Network Analysisand Mining14(1), 38 (2024).

[15]. Dassanayaka, S., Swed, O., & Volchenkov, D. (2024). Mapping the russian internet troll network on twitter using a predictive model. arXiv preprint arXiv:2409.08305.